\documentstyle[aps]{revtex}
\begin{document}
\twocolumn[\hsize\textwidth\columnwidth\hsize\csname
@twocolumnfalse\endcsname
\title{A comment on trans-Planckian physics in inflationary universe}
\author{Takahiro Tanaka}
\address{Yukawa Institute for Theoretical Physics, 
 Kyoto University, Kyoto 606-8502, Japan}
\maketitle

\thispagestyle{empty}
\begin{abstract}
There are several works searching for a clue of trans-Planckian 
physics on the primordial density perturbation spectrum. 
Here we would like to point out an important aspect which 
has been overlooked so far in this context. 
When we consider a model in which  
the primordial density perturbation spectrum is modified 
due to trans-Planckian physics, the energy density of fluctuations 
of the inflaton field necessarily becomes significantly large, 
and hence its back reaction to the cosmic 
expansion rate cannot be neglected. 
\vspace{5mm}
\end{abstract}]

\baselineskip10pt
In quantum field theory in curved space 
we usually decompose the field operator 
by using mode functions, which are 
the appropriately normalized 
solutions of the field equation. 
Many phenomena in early universe and 
radiation in black hole spacetime have been discussed 
based on this effective theory. 
In some cases, mode functions which had an infinitesimally short 
wave length are redshifted by an infinitely large amount, and 
become relevant modes which contribute to observable effects. 
In such cases, a simple question arises: 
Does any observable effect appear 
if we assume a certain modification 
for trans-Planckian physics? 
There are at least two cases in which context this question 
has been discussed.   
One is the Hawking radiation
\cite{Hawking,Unruh,Unruh1,JacCor,Corley,BMPS,CoJ,lattice,HiTa,SaSa} 
and the other is the quantum dynamics of a scalar field 
in an expanding universe, especially 
in the context of the inflationary universe scenario
scenario\cite{BrMa,Nie,JacMat,Kempf}. 
(For a review of the general issue as to trans-Planckian 
redshifts in cosmology and black hole physics, 
See Ref.\cite{YKIS} and references therein.)

In the former context, people studied models with 
a modified dispersion relation in most cases so far. 
Basically they reported that the thermal 
spectrum of Hawking radiation is reproduced in spite of 
introduction of a modified law of trans-Planckian physics. 
In the latter context, this issue  
is discussed only recently,  
in which they again examined models with a modified dispersion
relation. The main question in the latter studies is  
whether a modification of trans-Planckian physics can cause 
a non-scale invariant spectrum of primordial density fluctuations. 
However, there seems to be an aspect which has been overlooked
so far, but which is rather important in 
discussing trans-Planckian physics consistently in the context of 
the inflationary universe scenario. 

In the standard inflation scenario, the quantum state 
of the inflaton field is thought to be set almost to be in a ``vacuum'' 
initially at a time after the wavelength of each mode becomes 
sufficiently long compared with the Planck scale. 
As is known well, the concept of the vacuum state 
is quite ambiguous in curved space. In the context of the 
inflationary universe scenario, however, this ambiguity 
can be ignored by choosing the adiabatic vacuum 
when the wavelength of a mode is much shorter 
than the curvature scale of spacetime. 
This prescription applies in the standard inflationary 
universe scenario. 
Together with this assumption as to the initial condition, we also 
assume that the quantum field theory in curved space is 
valid after the wavelength becomes much longer than the 
Planck scale. As a result of these assumptions, 
we usually conclude the appearance of 
an almost scale invariant spectrum for initial fluctuations. 

Even when we assume a non-trivial trans-Planckian physics, 
we suppose that physics after the wavelength becomes much 
longer than the Planck scale can be described by the standard 
quantum field theory in curved space. If this is not the case, 
the modification should not be referred to that of trans-Planckian physics. 
Thus, if we discuss the evolution of each mode after 
its wavelength becomes much longer than the Planck scale, 
i.e., if we restrict our attention to the regime in which 
the standard quantum field 
theory in curved space applies, the effect of 
a modified Planck scale physics comes into play only through the  
initial condition. 
Hence, in order to construct a model which leads to 
results different from the standard prediction, 
the quantum state of the inflaton 
field must be chosen differently from the ordinary 
vacuum as the initial condition at the very beginning.  
In this sense, a non-trivial initial condition is necessarily 
assumed when we study trans-Planckian physics in 
the cosmological context.  

However, setting a non-trivial initial quantum state for the inflaton 
field means, in some sense, that we consider a situation in which 
there exist a significant amount of inflaton particles. 
Even if we restrict our consideration to 
the modes whose wavelength has already become much longer 
than the Planck length, there are a lot of particles which have 
a relatively high momentum in comparison with the Hubble scale. 
The energy density of these inflaton particles 
will be given by $\int dp\, p^3 n_p$ denoting the occupation number of 
the modes with momentum $p$ as $n_p$. As we will see later, 
$n_p$ cannot be much smaller than unity in order to obtain 
a significantly modified initial power spectrum. 
Since we assumed that the physics is modified only at the 
Planck scale $M_{pl}$, it will be natural to suppose that 
the cutoff scale of $p$ below which the energy density 
can be well approximated by the above expression 
is not much smaller than $M_{pl}$.  
Hence the cutoff scale is much greater than 
$\sqrt{M_{pl}H}$, where $H$ is the expansion rate 
of the universe.  Then we find that 
the energy density of these inflaton particles 
dominates over the vacuum energy density ($\approx M_{pl}^2 H^2$) 
which is supposed to drive the universe to expand.
Almost the same argument is given in references \cite{LidLyt} 
to discuss robustness of the primordial spectrum of perturbations. 

To make the above argument more transparent, we would like to present
a more explicit calculation briefly. 
For simplicity, we consider a flat de Sitter model,  
whose metric is given by 
\begin{equation}
 ds^2={1\over H^2\eta^2}\left(-d\eta^2+d{\bf x}^2\right).  
\end{equation}
We treat here fluctuations of the 
inflaton field as a massless scalar field. 
As usual, the fluctuation field will be decomposed by using mode 
functions $u_{\bf k}(x)$ as
\begin{equation}
 \phi=\int {d^3k\over (2\pi)^{3/2}}\left(a_{\bf k}u_{\bf k}(x)
         +a^{\dag}_{\bf k}u^{*}_{\bf k}(x)\right),
\end{equation}
where $a_{\bf k}$ and $a^{\dag}_{\bf k}$ are annihilation and 
creation operators, respectively. 
The natural adiabatic vacuum state, i.e., so-called Bunch-Davies 
vacuum state, 
is specified by $a_{\bf k}|0_a\rangle=0$ with 
\begin{equation}
 u_{\bf k}={H\eta\, e^{-ik\eta} \over \sqrt{2k}}
         \left(1-{i\over k\eta}\right)e^{i{\bf k}\cdot{\bf x}}. 
\end{equation}

Here we consider a situation in which the quantum state of 
fluctuations of the inflaton field is in another vacuum state 
specified by 
\begin{equation}
 b_{\bf k}\vert 0_b\rangle =0,
\end{equation}
with 
\begin{equation}
 \phi=\int {d^3k\over (2\pi)^{3/2}}\left(b_{\bf k}v_{\bf k}(x)
         +b^{\dag}_{\bf k}v^{*}_{\bf k}(x)\right),
\end{equation}
and
\begin{equation}
 v_{\bf k}=\alpha_{\bf k}u_{\bf k}+\beta_{\bf k}u_{\bf k}^{*}. 
\end{equation}

Assuming this $b$-vacuum state, we compute the amplitude of 
a fluctuation mode at a late epoch after its wavelength becomes 
much longer than the Horizon scale as  
\begin{equation}
 \langle 0_b|\,|\phi_{\bf k}|^2\,|0_b\rangle\,
\longrightarrow\hspace{-20pt}\lower-4pt\hbox{${}^{\eta\to 0}$}\,
  {H^2\over 2k^3}\left(|\alpha_{\bf k}|^2+|\beta_{\bf k}|^2
    +2|\alpha_{\bf k}|\cdot|\beta_{\bf k}|\cos\theta\right),
\end{equation}
where we have introduced 
$\theta=\arg(\alpha_{\bf k})-\arg(\beta_{\bf k})$.
Using the well-known relation $|\alpha_{\bf k}|^2=1+|\beta_{\bf k}|^2$, 
this amplitude of a fluctuation is found to be bounded from above as 
\begin{equation}
 \langle0_b|\, |\phi_{\bf k}|^2 \,|0_b\rangle < 
  {H^2\over 2k^3}\left(\sqrt{1+|\beta_{\bf k}|^2}+|\beta_{\bf k}|
    \right)^2. 
\end{equation}
From this, we can conclude that the occupation number 
$n_{\bf k}=|\beta_{\bf k}|^2$ cannot be much smaller than unity 
under the condition that 
$\langle0_b|\, |\phi_{\bf k}|^2 \,|0_b\rangle$ 
significantly differs from $H^2/2k^3$, the value 
corresponding to the Bunch-Davies vacuum.  

On the other hand, the expectation value of the 
energy momentum tensor, 
$T_{\mu\nu}=\phi_{;\mu}\phi_{;\nu}
      -{1\over 2}g_{\mu\nu}g^{\rho\sigma}\phi_{;\rho}\phi_{;\sigma}$, 
will be given by 
\begin{eqnarray}
\langle 0_b|\, (-T_0^0)\,|0_b\rangle 
   & = & {H^2\eta^2\over 2}\int {d^3 k\over (2\pi)^3}\cr
  &&
    \times \Biggl[(\alpha_{\bf k} u'_{\bf k}+
            \beta_{\bf k} u'_{\bf k}{}^{*})
           (\alpha_{\bf k}^{*} u'_{\bf k}{}^{*}+
            \beta_{\bf k}^{*} u'_{\bf k})\cr 
&&\quad +{\bf k}^2
           (\alpha_{\bf k} u_{\bf k}-
            \beta_{\bf k} u_{\bf k}^{*})
           (\alpha_{\bf k}^{*} u_{\bf k}^{*}-
            \beta_{\bf k}^{*} u_{\bf k})
           \Biggr] \hspace{-10mm}\cr
 &\longrightarrow\hspace{-28pt}\lower-4pt\hbox{${}^{\eta\to -\infty}$}& 
           {(H\eta)^4\over 2(2\pi)^3}
           \int d^3 k\cdot k\Biggl[|\alpha_{\bf k}|^2
           +|\beta_{\bf k}|^2 \cr &&
\qquad
          -(\alpha_{\bf k}\beta_{\bf k}^* e^{-2ik\eta}
           +\alpha_{\bf k}^* \beta_{\bf k} e^{2ik\eta}
           )\Biggr]. 
\label{above}
\end{eqnarray}
As is known well, this expression is divergent for any choice of 
$\alpha_{\bf k}$ and $\beta_{\bf k}$. To obtain a finite result, 
it is necessary to implement some renormalization procedure 
which subtract appropriate covariant counter terms 
written in terms of geometrical quantities\cite{BD}. 
Here we adopt a simple prescription in which we just subtract the 
expectation value of the energy momentum tensor for the Bunch-Davies 
vacuum, and we refer to it as a renormalized quantity. 
Namely, we calculate   
$\langle 0_b|\, (-T_0^0) \,|0_b\rangle^{(ren)}\equiv 
   \langle 0_b|\,  (-T_0^0) \,|0_b\rangle
    -\langle 0_a|\,  (-T_0^0) \,|0_a\rangle$. 
To remove the highly oscillatory contribution from the 
terms in the round brackets in the last line of Eq.(\ref{above}), 
we consider a time averaged quantity. Then we find 
that the energy density due to the inflaton particles is 
given by 
\begin{eqnarray}
{1\over \Delta\eta}\int_{\eta-\Delta\eta/2}^{\eta+\Delta\eta/2} d\eta'
   \langle (-T_0^0) \rangle^{(ren)}
  &\approx& {(H\eta)^4\over (2\pi)^3}
    \int d^3k\cdot k \vert\beta_{\bf k}\vert^2\cr
   &=&{1\over (2\pi)^3}
    \int d^3p\cdot p\,  n_p,  
\label{fin}
\end{eqnarray}
where we have introduced the physical momentum $p\equiv k/a$. 
Now we have obtained the anticipated result.

As we have mentioned earlier, 
this result tells us that the energy density due to fluctuations of 
the inflaton field becomes larger than that due to the inflaton 
potential as long as the cutoff scale for momentum below which 
the ordinary quantum field theory in curved space holds 
is chosen to be greater than $\sqrt{M_{pl} H}$ or 
equivalently $V^{1/4}$, where $V$ is the 
energy density due to the inflaton potential. 
When we discuss the effect of trans-Planckian physics, 
this cutoff scale would be supposed to be near the Planck scale. 
In such cases, the basic assumption as to the background 
inflationary universe will not be appropriate because the 
back reaction becomes significant. 
Here we mention that the order of magnitude of 
Eq.(\ref{fin}) does not change even if 
we consider the case in which $n_{\bf k}$ is non-trivial only 
for fluctuations in a rather narrow band of $k$.  

If we wish to construct a consistent model with a large value of 
$n_{\bf k}$, we need to invent a certain mechanism of generating
negative energy density which cancels a large amount of energy density due 
to fluctuations of the inflaton field. 
Moreover, even if we could construct a model with such a 
mechanism, we need to suppose that $n_{\bf k}$ is not constant in 
$k$ in order to realize a non-trivial initial power spectrum. 
Then, it is easy to see from Eq.(\ref{fin}) 
that the energy density due to fluctuations 
becomes time-dependent. 
Again, if we consider the case in which $n_{\bf k}$ is non-trivial only 
in a narrow band of $k$, 
the energy density due to fluctuations behaves as a radiation field, 
i.e, $\propto a^{-4}$.
Therefore, it is not clear whether 
the background accelerated expansion of the universe is realized. 

Furthermore, it seems that we also need to pay attention to 
the fact that non-trivial vacuum means existence of 
particles in some sense, although we did not discuss 
it in this short report. Under the circumstance filled with 
a lot of particles with high momentum, interaction between 
particles should be taken into account. 

\newpage
\centerline{\bf Acknowledgement}
The author thanks Ted Jacobson and Misao Sasaki for their 
valuable comments. He also acknowledges support from Monbusho 
Grant-in-Aid No. 1270154.

\end{document}